\documentclass[prd, twocolumn, nofootinbib, floatfix]{revtex4-2} 

\usepackage{amsmath}
\usepackage{graphicx}
\usepackage{dcolumn}
\usepackage{bm}
\usepackage{epsfig}
\usepackage{amssymb,latexsym,mathrsfs,empheq}
\usepackage{graphicx}
\graphicspath{{./}{figures/}}
\usepackage{color}
\usepackage{xcolor} 
\usepackage{mathtools} 
\usepackage{hyperref} 
\usepackage{orcidlink}

\hypersetup{
    colorlinks=true,
    linkcolor=red,
    citecolor=blue,
} 

\newcommand{\be}{\begin{equation}}
\newcommand{\ee}{\end{equation}}
\newcommand{\bs}{\begin{split}} 
\newcommand{\bea}{\begin{eqnarray}}
\newcommand{\eea}{\end{eqnarray}} 
\newcommand{\al}{\alpha}

\newcommand{\Om}{\Omega_m} 
 
\newcommand{\Or}{\Omega_r} 
 
\newcommand{\winf}{w_\infty} 
\newcommand{\vpi}{\varphi_f} 
\newcommand{\vpf}{\varphi_f} 
\newcommand{\astar}{a_\star}

\usepackage{soul}

\begin{document}

\title{Connecting Primordial Gravitational Waves and Dark Energy} 

\author{Tilek Zhumabek$^{1,2}$\orcidlink{0000-0001-7900-786X}, Mikhail Denissenya$^{1}$\orcidlink{0000-0003-4734-7127}, Eric V.\ Linder$^{1,3}$\orcidlink{0000-0001-5536-9241}} 
\affiliation{
$^{1}$Energetic Cosmos Laboratory, Nazarbayev University, 
Astana 010000, Qazaqstan\\ 
$^{2}$Department of Physics, School of Sciences and Humanities, Nazarbayev University, 
Astana 010000, Qazaqstan\\
$^{3}$Berkeley Center for Cosmological Physics \& Berkeley Lab, 
University of California, Berkeley, CA 94720, USA
} 

\begin{abstract}
Cosmic acceleration manifested in the early universe as 
inflation, generating primordial gravitational waves detectable in 
the cosmic microwave background (CMB) radiation. Cosmic 
acceleration is occurring again at present as dark energy, 
detectable in cosmic distance and structure surveys. We explore the 
intriguing idea of connecting the two occurrences through 
quintessential inflation by an $\al$-attractor potential 
without a cosmological constant. For this model we demonstrate robustness 
of the connection $1+w_0\approx 4/(3N^2r)$ between the present day 
dark energy equation of state parameter $w_0$ and the primordial 
tensor to scalar ratio $r$ for a wide range of initial conditions. 
Analytic and numerical solutions produce  
current thawing behavior, resulting in a tight relation  
$w_a\approx-1.53(1+w_0)\approx -0.2\,(4\times 10^{-3}/r)$. 
Upcoming CMB and galaxy redshift surveys 
can test this consistency condition. Within this 
model, lack of detection of a dark energy deviation from 
$\Lambda$ predicts a higher $r$, and lack of detection of $r$ 
predicts greater dark energy dynamics. 
\end{abstract} 

\date{\today} 

\maketitle

%%%%%%%%%%%%%%%%%%%%%%%%%%%%% 
\section{Introduction} 

Inflation and dark energy govern two major epochs of cosmic 
history and drastically change the universe. Inflation, 
through quantum fluctuations, generates structure from the 
vacuum, creating both matter and radiation density perturbations 
and tensor perturbations of spacetime itself, that are 
primordial gravitational waves. Dark energy has come to 
dominate the cosmic energy density recently, leading to rapid 
expansion of distances and suppression of the growth of cosmic 
structure that occurred during the matter dominated epoch. 

While accelerated expansion is in common between the 
two eras, they are separated by billions of year in time, or 
some 60 e-folds in expansion. Thus the idea of unifying their 
mechanisms, known as quintessential inflation (see, e.g., 
\cite{9810509,9811385,0111417,0504191,1410.6100,2106.14966,2108.11144,2112.11948}), 
is both attractive and problematic. If they arise from the 
dynamics of a scalar field rolling on its potential, the 
potential itself is difficult to rationalize and the difference 
in energy scales (roughly $10^{25}$ eV for inflation, 
$10^{-3}$ eV for dark energy) is large. The hierarchy problem 
is well known and is a general puzzle even if we ascribe 
dark energy simply to a cosmological constant \cite{wbg}. 
However ideas from supergravity present a robust concept 
for the inflation potential through $\alpha$-attractors 
\cite{1306.5220,1311.0472,1412.3797}. 
As well $\al$-attractors can be 
employed for dark energy, e.g.\ \cite{1505.00815,2005.14053}. Finally, the 
unification of the two as quintessential inflation has been 
considered, e.g.\ \cite{1703.00305,1712.01760,1712.09693,1803.00661,2010.15822,2103.07892,2106.14966,2305.15378}. 

We explore $\al$-attractors, which includes Starobinsky 
gravity, as quintessential inflation in more detail, 
demonstrating and highlighting the important characteristic that they behave 
as thawing scalar fields. This helps motivate the large gap 
in time between the two episodes of cosmic acceleration. 
The field rolled along its plateau during inflation, ending 
when the potential steepened and the energy density diminished, 
but Hubble friction governed the dynamics, freezing the 
field until recently when 
the field was released to thaw and gradually deviate from 
its cosmological constant-like torpor. Thawing fields have 
specific characteristics that can be derived analytically, and 
the equations of motion can be evaluated numerically to check. 

The remarkable aspect of quintessential inflation combined 
with the $\al$-attractor formalism is that predictions from 
the two eras are tightly connected, moreover in an ``everybody wins 
something'' manner. If the tensor to scalar ratio $r$ of perturbation 
power generated by inflation is low (making it difficult 
for future CMB experiments to detect), i.e.\ the potential is 
too flat, then it freezes higher up its potential and has 
more dynamics once it thaws in the dark energy phase. Conversely, 
if $r$ is large enough for easier detection due to the potential 
being steeper then the field traverses further and thaws recently 
in a flatter part of the potential, giving less dark energy dynamics. 
We explore which wins under what circumstances, and whether 
there is a happy medium where both experimental signals 
are accessible. We work with a potential with no cosmological 
constant, i.e.\ zero minimum, so that all the cosmic acceleration 
is from the $\al$-attractor dynamics, and that none was ``put 
in by hand''. 

Section~\ref{sec:alatt} provides a brief summary of 
$\al$-attractors and sets up the coupled system of 
equations of motion. In Section~\ref{sec:analysis} we 
solve the dynamics, demonstrating the dependence of results 
on physics inputs such as $\al$ and the frozen field value. 
We demonstrate in Section~\ref{sec:scan} both 
analytically and numerically that 
viable results belong to the thawing class of dark energy 
and have definite relations between $r$ and the dark energy 
equation of state parameters $w_0$ and $w_a$. 
Section~\ref{sec:expt} discusses constraints from future 
CMB and cosmic distance and structure experiments, with results from 
each guiding the other. We discuss and summarize our results 
in Section~\ref{sec:concl}.

%%%%%%%%%%%%%%%%%%%%%% 
\section{$\alpha$-Attractors as Quintessential Inflation} \label{sec:alatt} 

The class of $\al$-attractor models has several attractive 
features, including high energy physics and symmetry 
motivations; see \cite{1306.5220,1311.0472,1412.3797} for details. These involve a 
Lagrangian density with a scalar field with a pole kinetic 
term and a potential, 
\be 
\mathcal{L}= \frac{1}{2}R - \frac{1}{2}\frac{ (\partial \phi)^2}{\left(1-\frac{\phi^2}{6\alpha}\right)^2} - V\left(\frac{\phi}{\sqrt{6\alpha}}\right)+\mathcal{L}_m\ ,  
\ee 
where $\mathcal{L}_m$ is the matter Lagrangian and we 
set the reduced Planck mass $M_P=8\pi G=1$ (so $\phi$ 
is dimensionless). 
The field has poles at $\phi=\pm\sqrt{6\alpha}$. The 
parameter $\alpha$ controls the position of the pole 
\cite{1405.3646},  
and can also be interpreted 
geometrically in taking discrete values related to the 
symmetry class \cite{1610.04163,1704.04829}. 

Due to the behavior near the pole, the field rolls very 
slowly, allowing for a high energy inflationary epoch. 
Quite generically this leads to predictions for the 
inflationary curvature perturbation power spectrum tilt 
$n_s$ and the tensor to scalar perturbation power ratio 
$r$, 
\be 
n_s = 1-\frac{2}{N}\ , \quad r  =\frac{12\alpha}{N^2} \ , 
\label{eq:rns}
\ee 
where $N$ is the number of e-folds between horizon crossing 
and the end of inflation, generally of order 50--60. 
The Poincar\'e disk scenario of $\al$-attractors gives 
discrete values of $3\al=1,2,\dots 7$, while Starobinsky 
and Higgs inflation models possess $\al=1$. 

Specific details of the inflationary evolution will 
depend on the potential chosen. It is convenient to use 
a canonically normalized field $\varphi$ instead of the 
original field $\phi$ through the transformation 
\be 
\phi = \sqrt{6\alpha}\, \tanh{\frac{\varphi}{\sqrt{6\alpha}}}\ . 
\ee 
This moves the poles to $\varphi=\pm\infty$ and explicitly 
demonstrates the inflationary plateau where the field 
rolls slowly. The potential $V(\varphi)$ generally rolls  
off the plateau at large $\varphi$ with an exponential 
deviation, due to the tanh function. 

We are particularly interested in quintessential inflation 
$\al$-attractor models, where after inflation the field rolls 
to much smaller values (by $\sim10^{110}$) of the potential 
energy suitable for late time dark energy cosmic acceleration. 
One way of achieving late time acceleration is including a 
cosmological constant $\Lambda$ in the potential, i.e.\ another 
plateau at energy scale $\Lambda$. We do not take this 
approach, but rather seek a dynamical dark energy with 
zero cosmological constant. Furthermore, just as there is 
attractor behavior at early times regarding the observables 
$n_s$ and $r$, we want to avoid fine tuning by using a 
potential with attractor behavior at late times, of an 
exponential potential form. 

These desiderata are satisfied by the potential, 
\bea  
V(\phi) &=& M^2 e^{-2g}\,\left[e^{g\left(\frac{\phi}{\sqrt{6\alpha}} + 1\right)} -1\right]\\ 
V(\varphi) &=& M^2 e^{-2g}\,\left[e^{g\left(\tanh(\varphi/\sqrt{6\al}) + 1\right)} -1\right]\ ,  
\eea  
called by \cite{1712.09693} as Exp-model II; we call it the 
ExpLin model since $V(\phi)$ approaches the positive 
pole as an exponential (giving inflation) and the negative 
pole as a linear function in $\phi+\sqrt{6\al}$ (which 
translates to an exponential in $\varphi$; note the linear potential, in 
addition to the pole structure,  
gives some protection against quantum corrections for this dark 
energy part of the potential). 
Here $M^2$ gives the inflation energy scale and 
$M^2 e^{-2g}$ will be of order the current dark energy 
density (hence $g\approx125$; while still not of order 
one, the exponential relaxes the huge hierarchy). 

Since the dark energy part of the potential has an 
exponential form, 
\be 
V(\varphi\to -\infty)\approx M^2 e^{-2g}\,2g\,e^{-2|\varphi|/\sqrt{6\al}}\ , \label{eq:lateV} 
\ee 
i.e.\ $V\sim e^{-\lambda\varphi}$, then 
we will have an attractor behavior 
to $\winf=-1+\lambda^2/3=-1+2/(9\al)$. 
Thus we have the very nice relation that 
\be 
1+\winf=\frac{8}{3N^2r}=0.22\,\left(\frac{51}{N}\right)^2\left(\frac{4.6\times10^{-3}}{r}\right)\ . \label{eq:winfr}
\ee 
(The parameter values used for illustration correspond to the simple Starobinsky model as given in 
\cite{2202.02773}.) 
That is, if  $r$ is so small as to evade detection 
in CMB polarization B-modes, we stand a good chance 
of seeing a signature in late time acceleration different 
from a cosmological constant ($w=-1$), while if there 
is no deviation from a cosmological constant observed 
then this predicts (within this quintessential inflation 
model) that inflationary gravitational waves should be 
detected at a reasonable value of $r$. 

We solve the exact dynamics of the quintessential 
inflation field through numerical evaluation of its 
equation of motion, the Klein-Gordon equation, plus 
the Friedmann equation. 
We implement this as an autonomous system of coupled 
ordinary differential equations (see, e.g., \cite{copeland2006dynamics}) 
for the kinetic and potential energies, using variables 
$x = \varphi'/\sqrt{6}$, $y = \sqrt{V(\varphi)/(3H^2)}$: 
\bea
    x' &=& -3x +\sqrt{\frac{3}{2}}\,\lambda y^2 + \frac{3}{2}x\left[2x^2+\gamma(1-x^2-y^2)\right]\\
    y' &=&-\sqrt{\frac{3}{2}}\, \lambda y x+ \frac{3}{2} y\left[2 x^{2}+\gamma\left(1-x^{2}-y^{2}\right) \right]\ . 
\eea 
Here $H$ is the Hubble parameter and a prime denotes a 
derivative with respect to the e-fold time variable $\ln a$. 
The background equation of state parameter $\gamma$ 
(neglecting the field) and the 
fractional potential slope $\lambda$ are 
\bea 
\gamma &=& 1 + \frac{1}{3}\,\frac{1}{1+\frac{\Om}{\Or}\, e^{\ln a}}\\ 
\lambda &=& -\frac{V_{,\varphi}}{V}\ , 
\eea 
where $\Om$ and $\Or$ are the present fractions of energy 
density in matter and radiation, respectively, relative 
to the critical density. 

To test the numerical accuracy we have also implemented 
the coupled equations in the three equation 
form \cite{copeland2006dynamics}, defining 
$\Gamma = V V_{,\varphi \varphi}/V^2_{,\varphi}$ and 
including the evolution equation 
\be
\lambda' = -\sqrt{6}\,\lambda^2(\Gamma -1)x\ . 
\ee
The results are identical to desired precision, and we 
use the two equation system for all results presented.

%%%%%%%%%%%%%%%%%%%%%% 
\section{Dynamical Results} \label{sec:analysis} 

The field evolution is determined by the model 
parameter $\al$, and the initial conditions on the 
field, $\varphi_i=\vpf$ i.e.\ the frozen value. Note 
though that we expect the attractor behavior 
to give evolution substantially independent of $\vpi$ over 
some range. 
We set initial conditions at $\ln a=-15$ ($z\approx3\times 10^6$). 
Due to the high Hubble friction at early 
times, results are independent of $\varphi'_i$, and we 
set it to zero as appropriate for the frozen field state. 
The matter density today will affect late 
time results and our fiducial value is $\Om=0.3$ 
(and where necessary we adopt $H_0=68$ km/s/Mpc). 
We generally consider a range of $\al=[1/3,7/3]$, to 
match the values corresponding to the Poincar\'e disk 
$\al$-attractor origin. 

\begin{figure*}[ht]
\centering
 \includegraphics[width=0.48\textwidth]{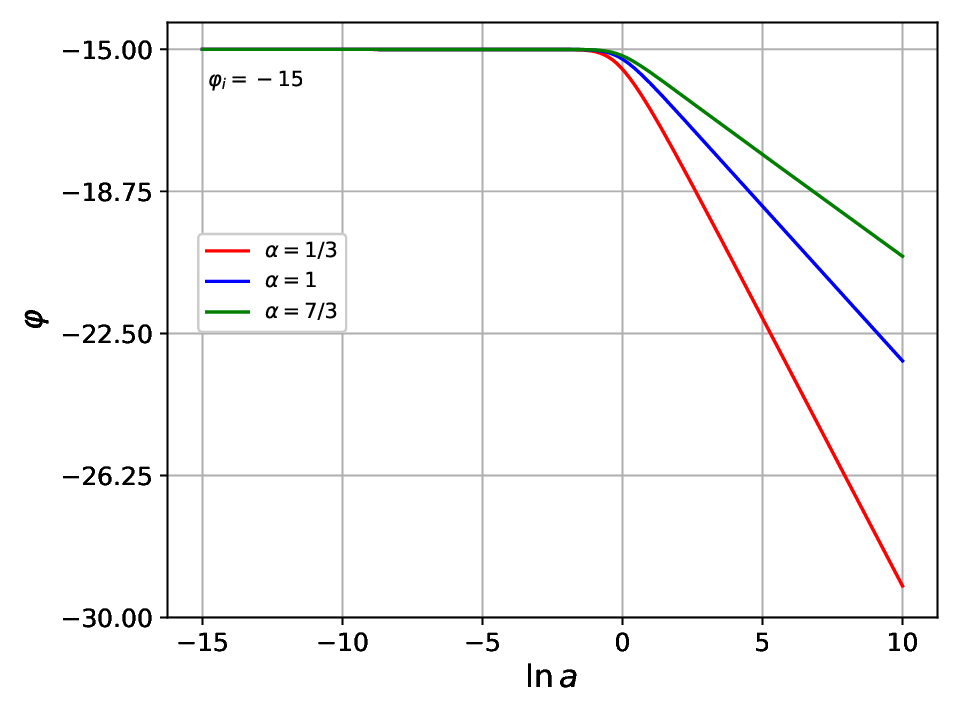}\quad 
        \includegraphics[width=0.48\textwidth]{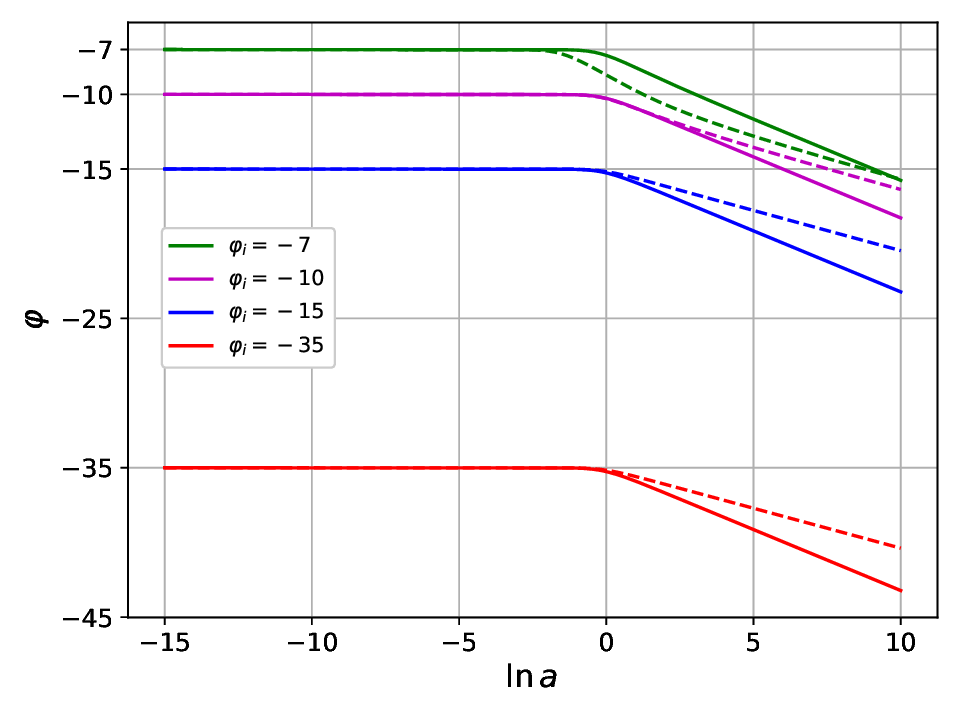}
        \caption{The field thaws from the frozen state induced by high Hubble friction, and rolls down the potential with a 
        rapidity determined by its steepness. [Left panel] 
        We illustrate the dependence on $\al$, fixing $\varphi_i\equiv\vpi=-15$. 
        [Right panel] We illustrate the dependence on $\vpi$, 
        fixing $\al=1$ (solid curves) or $\alpha=7/3$ (dashed curves). 
        } 
    \label{fig:phiva}
\end{figure*}

Figure~\ref{fig:phiva} shows the field evolution as a 
function of $\al$ and of $\vpi$. As expected, the field 
is frozen during the matter dominated epoch and only 
rolls at close to the present ($\ln a=0$). Such behavior is 
known as a thawing field \cite{0505494,0601052,1505.00815} and 
will have important physical consequences discussed in 
Sec.~\ref{sec:scan}. The smaller $\al$, the steeper the 
potential and the sooner and more vigorously it thaws 
(see the left panel). Also, the closer $\vpi$ is to zero, 
the steeper the potential (i.e.\ the further from the 
$\varphi\to -\infty$ approach to $V=0$) and 
again the field thaws sooner and more strongly (see the 
right panel).

%%%%%%%%%%%%%%%%%%%%%% 

\begin{figure*}[ht]
\centering
\includegraphics[width=0.48\textwidth] 
{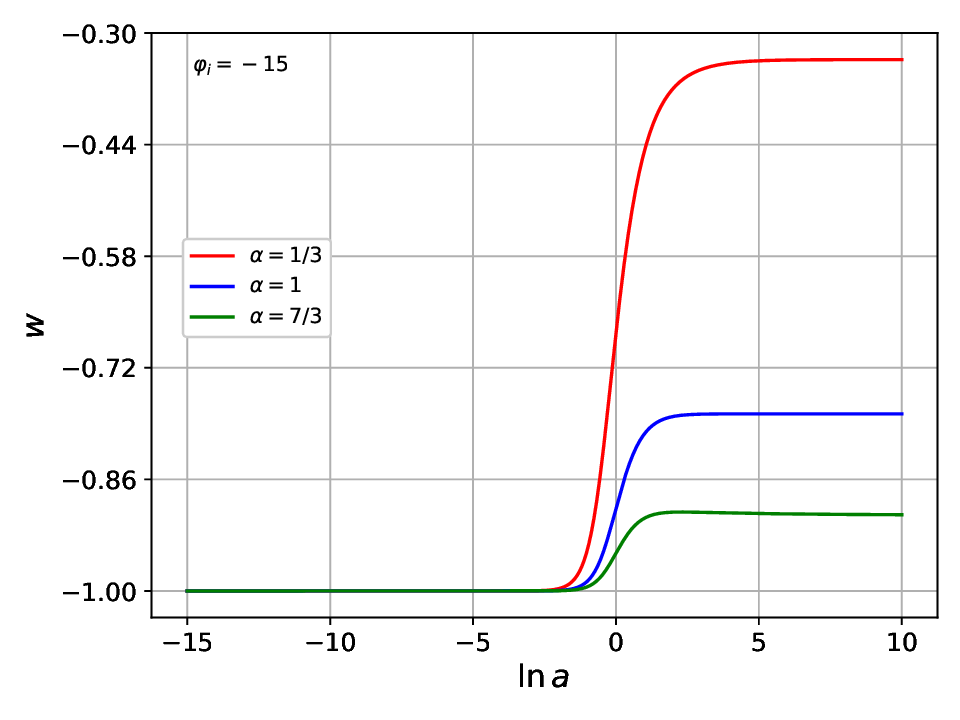}\quad 
 \includegraphics[width=0.48\textwidth] 
 {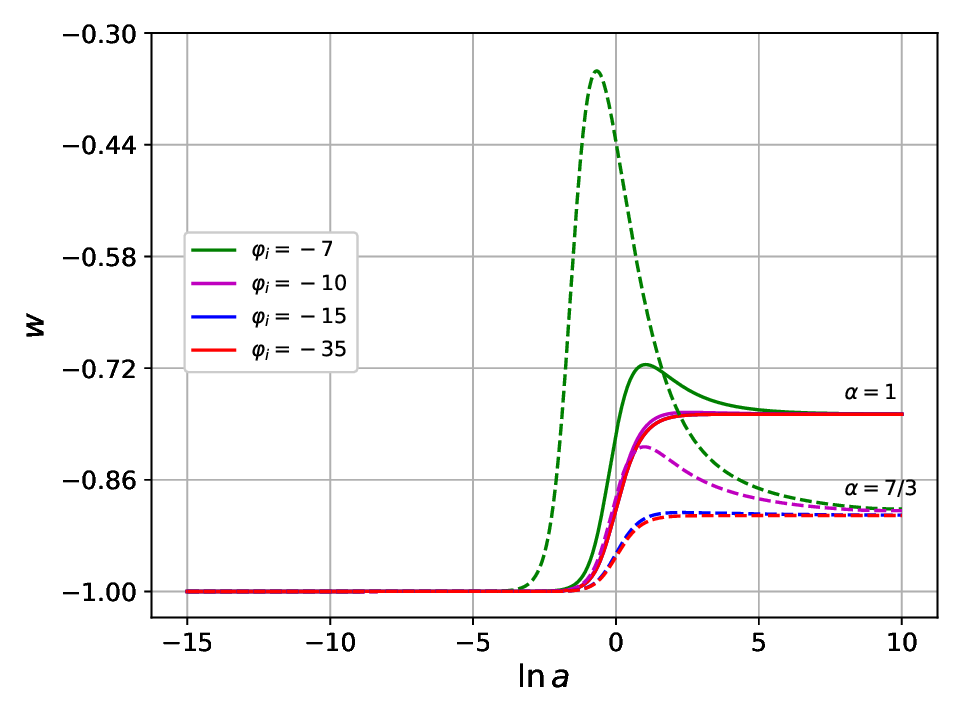} 
\caption{The dark energy equation of state parameter $w(a)$ 
evolves from a cosmological constant state to its future 
attractor value. [Left panel] The smaller $\al$ and hence 
the steeper the potential, the sooner and more that $w(a)$ 
deviates from $-1$. Values $\al<1$ will have difficulty 
being consistent with observations. 
[Right panel] The less negative $\vpi$ is and hence the 
steeper the potential, the sooner and more that $w(a)$ 
deviates from $-1$. Note that very steep potentials can cause 
$w$ to overshoot its attractor value. Values $\vpi>-10$ will 
have difficulty being consistent with observations, while 
values $\vpi<-15$ give near identical results. 
} 
    \label{fig:wvsa}
\end{figure*}

The dark energy dynamics can be usefully described through 
the equation of state parameter, or pressure to density ratio, 
$w(a)$. Figure~\ref{fig:wvsa} illustrates the evolutionary 
behavior. While the field is frozen, $w=-1$, and then as it 
thaws it moves away from cosmological constant behavior and 
becomes less negative. In the future it goes to its attractor 
behavior $\winf=-1+2/(9\al)$. Depending on the steepness of 
the potential it may overshoot and then relax to the attractor. 
Note that for $\vpi\lesssim-15$ the equation of state is rather 
insensitive to $\vpi$, while for $\vpi\gtrsim-10$ the deviation 
near the present is so extreme that it will be severely 
constrained by observations.

%%%%%%%%%%%%%%%%%%%%%%%%%%%%%%%%%  

%%%%%%%%%%%%%%%%%%%%%% 
\section{Relations between Inflation and Dark Energy} \label{sec:scan} 

As we have seen, the ExpLin $\al$-attractor model has 
a definite relation between the inflationary tensor to 
scalar ratio $r$ and the late time attractor dark energy 
equation of state parameter $\winf$. However, late time 
observations do not constrain $\winf$ directly but rather 
cosmic distances and growth factors that involve $w(a)$. 
Therefore we need to examine how $w(a)$ behaves. 

The form $w(a)=w_0+w_a(1-a)$ has been demonstrated 
\cite{0208512,0808.0189} to be an excellent fit to 
observations (distances and growth factors) relative to the 
exact equation of motion, to $\sim0.1\%$ accuracy, 
especially for thawing fields. Therefore we investigate 
the relation of the parameters $w_0$ and $w_a$ to the 
model parameters. 

\begin{figure*}[ht]
\centering 
        \includegraphics[width=0.48\textwidth]{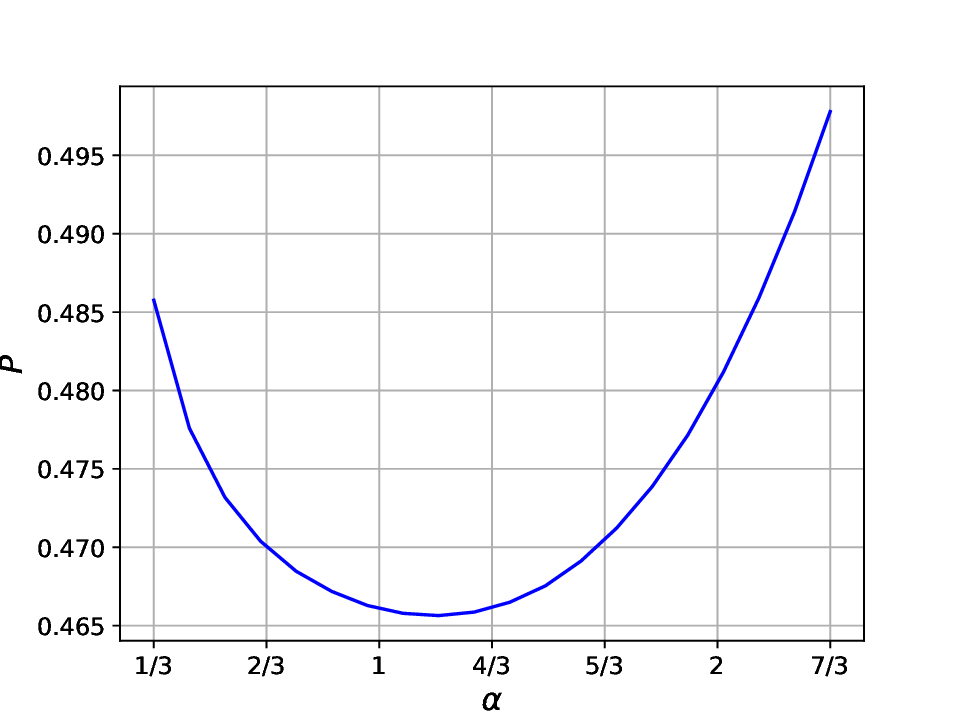}\quad 
        \includegraphics[width=0.48\textwidth]{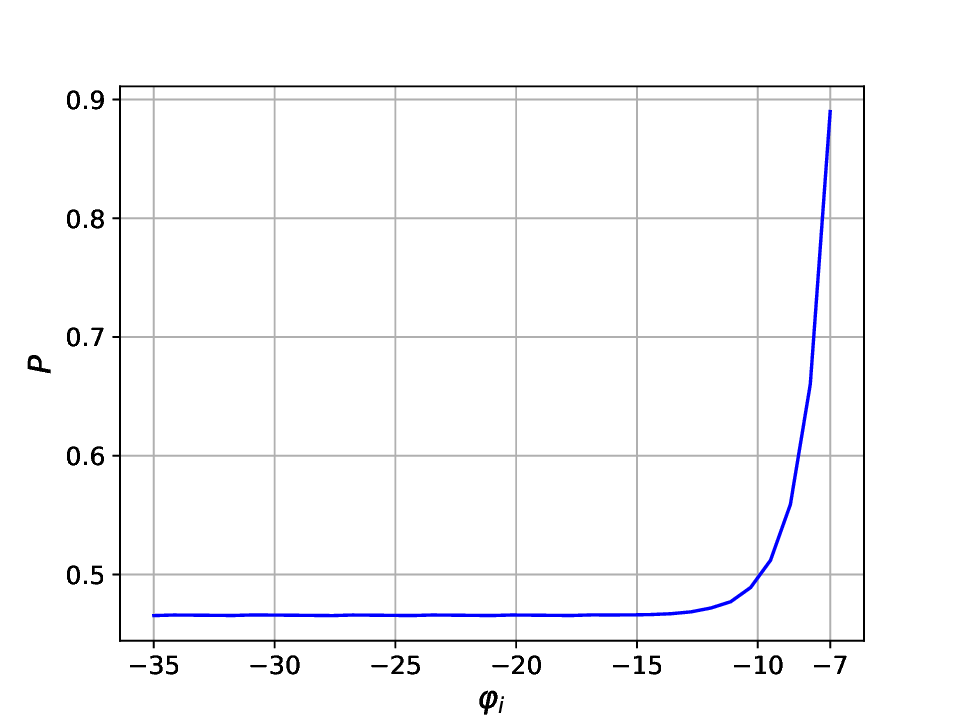} 
\caption{The ratio $P\equiv(1+w_0)/(1+\winf)$ is 
nearly constant over the range of $\al$ (left panel, 
with $\vpi=-15$) and viable 
$\vpi$ (right panel, with $\al=1$) considered. 
}
    \label{fig:pw0}
\end{figure*}

First, we connect $\winf(\alpha)$ to $w_0$. 
Figure~\ref{fig:wvsa} already hints that there is a close 
connection for viable models, i.e.\ those that do not 
overshoot the attractor to larger $w$. We quantify the 
ratio $P\equiv(1+w_0)/(1+\winf)$ in Figure~\ref{fig:pw0}. 
Indeed we find that over the range of $\al$ under 
consideration, and for viable $\vpi$, we have 
\be 
1+w_0\approx 0.5(1+\winf)\ , 
\ee 
to a good approximation. 
Thus we can relate inflationary 
$r$ to dark energy $w_0$, so Eq.~\eqref{eq:winfr} becomes 
\be 
1+w_0\approx\frac{4}{3N^2r}=0.10\,\left(\frac{51}{N}\right)^2\left(\frac{4.6\times10^{-3}}{r}\right)\ . \label{eq:w0r}
\ee

%%%%%%%%%%%%%%%%%%%%%%%%%%%%%% 

As for the time variation of the dark energy equation of state 
parametrized by $w_a$, we recall that many thawing fields 
have a narrow relation $w_a\approx -(1+w_0)\times(1.5-1.6)$ as demonstrated 
in \cite{0808.0189}. To investigate whether that holds for 
ExpLin we solve numerically the coupled equations of motion; 
note that $w(a)=(x^2-y^2)/(x^2+y^2)$. The parameter $w_0$ is 
simply $w(a=1)$, while Ref.~\cite{0808.0189} established 
that $w_a$ has a physical interpretation 
as a calibration, or stretching, parameter to unify members of 
a particular class (i.e.\ parameter values within a model). 
From the phase space 
quantity $w'$ we can define $w_a=-w'(a)/a$ and evaluate it at some $a=a_\star$ as described in \cite{0808.0189}. This only affects the theory {\it interpretation\/} -- bringing members of a theory family close together -- and not the observables that treat $w_0$ and $w_a$ as fit parameters within $w(a)=w_0+w_a(1-a)$. Following Ref.~\cite{0808.0189}, we seek to select $a_\star$ by where the phase space evolution $w$--$w'$ has a nearly universal track for different model parameters. 

\begin{figure*}[ht]
\centering 
\includegraphics[width=0.48\textwidth]{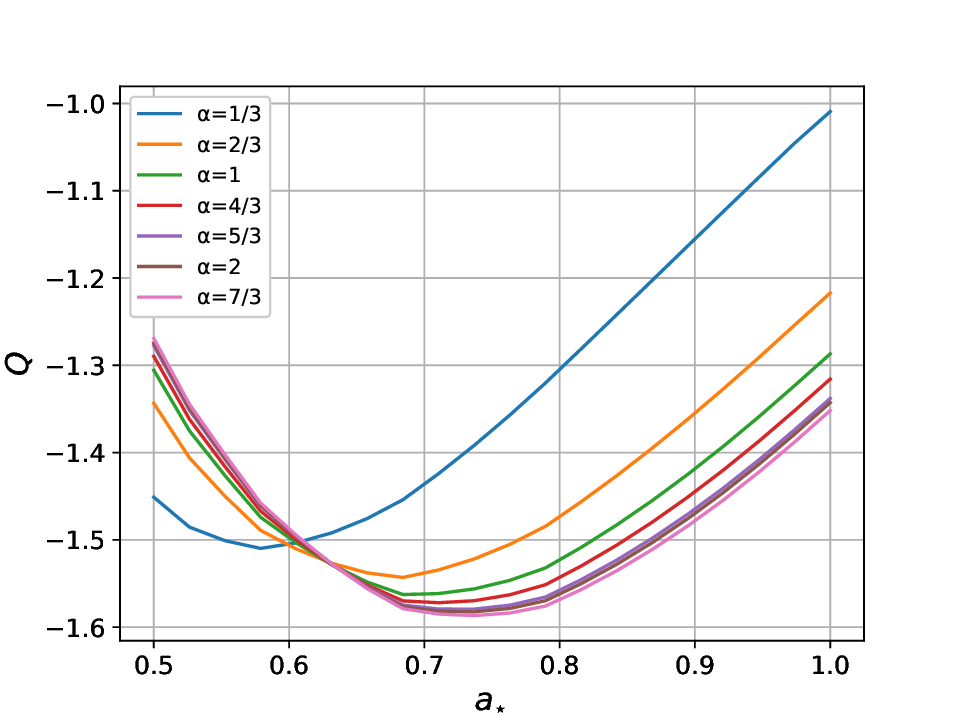}\quad 
        \includegraphics[width=0.48\textwidth]{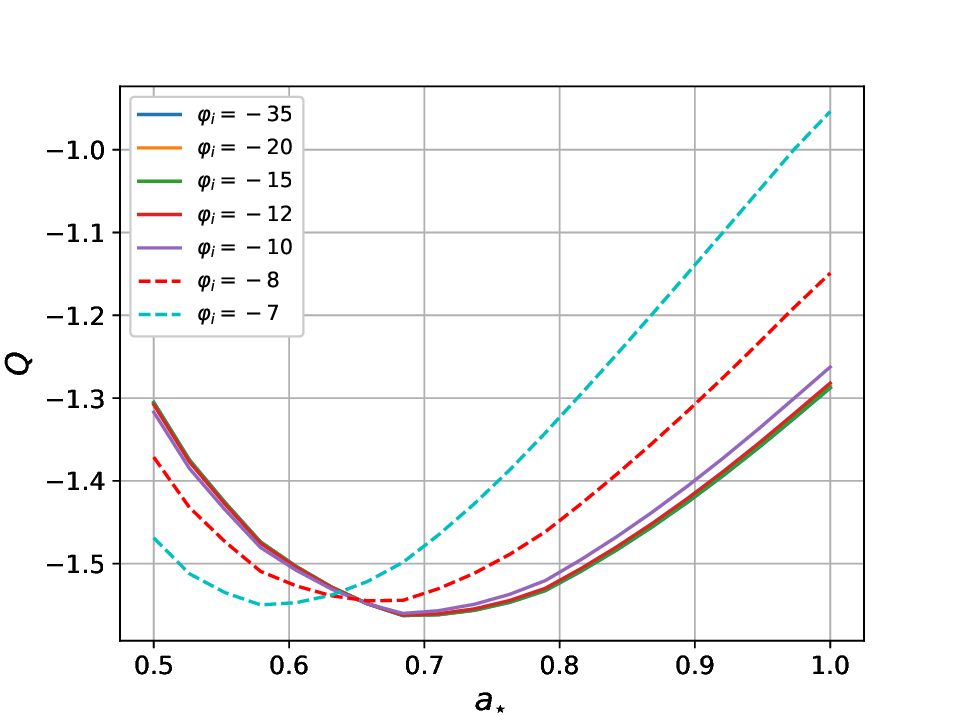} 
\caption{The ratio $Q\equiv w_a/(1+w_0)$, with $w_a=-w'(\astar)/\astar$, 
shows that the ExpLin model acts like a standard thawing field 
and is well calibrated (nearly universal phase space behavior -- 
curves at nearly the same value) for $\astar\approx0.65$ over 
the viable parameter range. [Left panel] Variation with respect 
to $\al$, for $\vpi=-15$; 
[Right panel] Variation with respect 
to $\vpi$, for $\al=1$. 
}
    \label{fig:qwa}
\end{figure*}

Figure~\ref{fig:qwa} demonstrates the calibration by computing 
$Q\equiv w_a/(1+w_0)$ as a function of $a_\star$. We see that 
most of the models lie close to each other for a broad range of 
$\astar$, and lie within the standard thawing behavior.  Both for 
variation of $\al$ (roughly $\alpha\gtrsim 2/3$: we can see from the left panel of Fig.~\ref{fig:wvsa} that $\alpha=1/3$ has a quite strong deviation from $w=-1$ and is not viable observationally) and $\vpi$ over its 
viable range the value $\astar\approx0.65$ (i.e.\ $z\approx0.5$) 
gives an especially tight calibration, i.e.\ 
\be 
w_a\approx -1.53(1+w_0) \ . \label{eq:waw0} 
\ee 
Thus we can also rewrite the relation Eq.~\eqref{eq:w0r} 
between inflation 
and dark energy as 
\be 
w_a\approx \frac{-6}{3N^2r}=-0.16\,\left(\frac{51}{N}\right)^2\left(\frac{4.6\times10^{-3}}{r}\right)\ . \label{eq:war}
\ee

%%%%%%%%%%%%%%%%%%%%% 

\begin{figure}[ht]
\centering 
\includegraphics[width=\columnwidth]{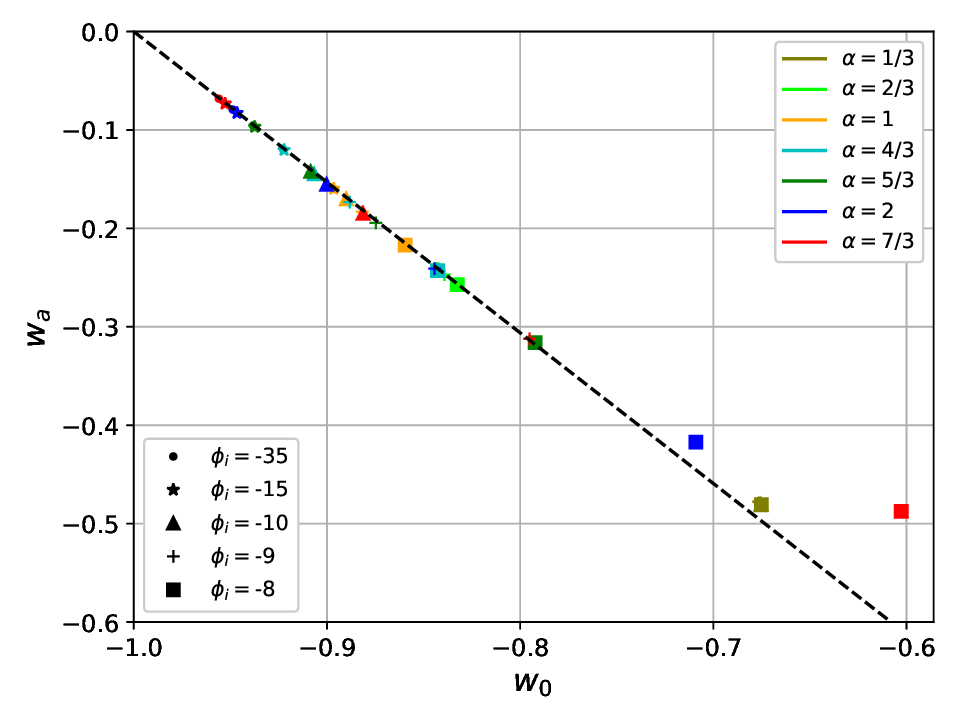} 
\caption{
Numerical solutions for dark energy equation of state 
parameters, denoted by colors and symbols corresponding 
to different $\alpha$ and $\vpi$, are in good agreement 
with Eq.~\eqref{eq:waw0} shown by the dashed line. 
Only outside of the viable region do deviations start 
to appear (note $\alpha = 1/3$ has all symbols on top of each other). 
}
    \label{fig:w0wa}
\end{figure}

To verify and illustrate the relations among 
$w_0$--$w_a$--$r$ we show the exact solutions for these 
quantities over a range of $\al$ and $\phi_i$. 
Figure~\ref{fig:w0wa} shows where the models lie in 
the $w_0$--$w_a$ space, with colors corresponding to 
different $\al$ and symbols to different $\phi_i$. They 
lie tightly clustered along the relation of Eq.~\eqref{eq:waw0}, 
showing that $w_a$ is indeed a dynamical physics calibration 
parameter. 
Slight deviations start to arise only outside the 
viable region, $w_0>-0.8$ (see Section~\ref{sec:expt}). 
All models depend little on $\vpi$ for $\vpi\lesssim-10$, 
and even out to $\vpi\approx-8$ stay on the $w_0$--$w_a$ 
relation; those that deviate at $\vpi>-8$ again tend to 
lie in the unviable regime.

%%%%%%%%%%%%%%%%%%%%%%%%%%  

\begin{figure*}[ht]
\centering 
\includegraphics[width=0.48\textwidth]{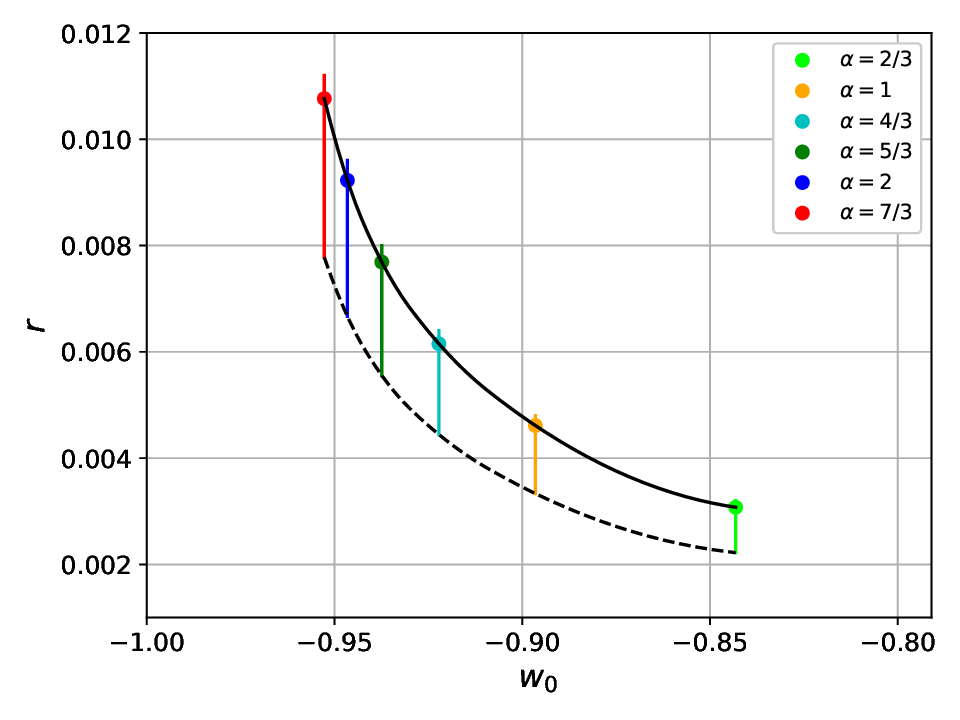}\quad 
        \includegraphics[width=0.48\textwidth]{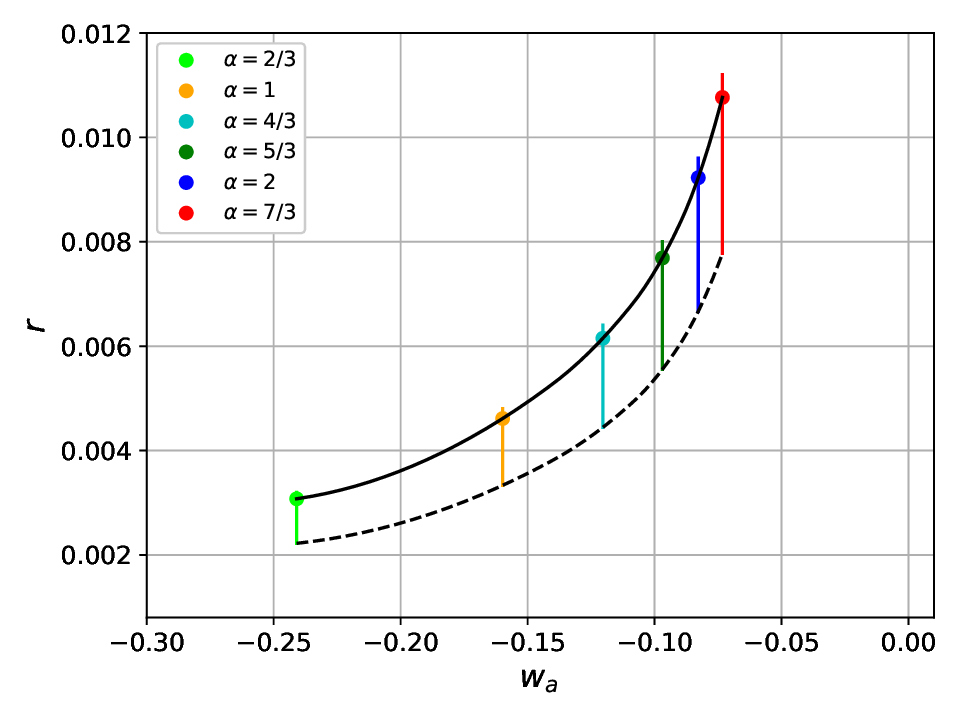} 
\caption{ 
The connection between the strength of inflationary  
gravitational waves, given by $r$, and the dark 
energy dynamics, given by $w_0$ (left panel) or 
$w_a$ (right panel), is tight as shown by the curves 
(here for $\vpi=-15$ but nearly the same over its 
viable range). 
Solid (dashed) black curves link the different $\alpha$ 
values for $N = 51$ (60), with vertical colored lines 
running over $N=[50,60]$ for constant $\al$. 
} 
    \label{fig:rw}
\end{figure*}

Figure~\ref{fig:rw} demonstrates the correspondence between 
inflationary $r$ and the dark energy parameters. For a given 
$\al$ the derived values for $r$, $w_0$, $w_a$ are plotted, 
using $\vpi=-15$ (but we just saw that $w_0$ and $w_a$ are 
rather insensitive to $\vpi$ over the viable range). The tensor to scalar ratio $r$ in Eq.~\eqref{eq:rns} scales as $1/N^2$ for fixed $\alpha$ (while dark energy dynamics does not depend on $N$, thawing from a late time frozen state), and we show a band from $N = [50,60]$ 
(this band should cover a reasonable range of inflation reheating scenarios), with the solid curve linking the different $\alpha$ values for $N = 51$. 
We explicitly 
see that dark energy dynamics and the strength of the primordial 
gravitational wave signature are inversely proportional, so that 
within this model if we fail to detect primordial gravitational 
waves we are likely to see dark energy dynamics distinct from 
a cosmological constant, while conversely if we fail to detect 
dark energy dynamics we are likely to detect primordial 
gravitational waves. With next generation experiments in 
both the CMB and cosmic distance/structure surveys, we 
may well detect both as measurement uncertainties should 
be of order (e.g.\ \cite{1611.00036,1809.01669,1907.04473,1910.09273}) 
$\sigma(r)\approx 5\times10^{-4}$, 
$\sigma(w_0)\approx0.08$, $\sigma(w_a)\approx0.2$ (while 
there are plans for experiments to improve these further).

%%%%%%%%%%%%%%%%%%%%%%%%  

%%%%%%%%%%%%%%%%%%%%%%%%%% 
\section{Constraints from Observations} \label{sec:expt} 

We can connect the quintessential inflation parameters 
to observables through fairly simple relations. First, 
from Eq.~\eqref{eq:rns} we have 
\bea 
N&=&\frac{2}{1-n_s}=57\,\frac{1-0.965}{1-n_s}\\ 
\al&=& \frac{r}{3(1-n_s)^2}=1.24\,\frac{r}{4.6\times 10^{-3}}\,\left(\frac{1-0.965}{1-n_s}\right)^2\ , 
\eea 
where $n_s$ and $r$ are observable from the CMB. 

The mass scale $M$ of the potential can be related to 
the amplitude of observed CMB temperature perturbations $A_s$ 
arising from inflation 
by (see, e.g., \cite{1712.09693})  
\be 
M^2=\frac{144\pi^2\al N}{(2N-3\al)^3}\,A_s\approx 10^{-10}\,\al\ . 
\ee 
Finally the exponential potential index $g$ comes from 
the late time dark energy epoch (see Eq.~\ref{eq:lateV}) by 
making sure the fractional dark energy today takes a value 
$\Omega_{{\rm de},0}=1-\Omega_m$ as desired. The parameter 
$g$ can efficiently be found for a particular 
value, e.g.\ $\Omega_{{\rm de},0}=0.7$, by bisection. 
Results for $g$ ($\approx 120$--128) 
as a function of $M^2$ and 
$\vpi$ are shown in the 
top right panel of Fig.~15 of \cite{1712.09693} 
(note they write $\gamma$ instead of $g$ and $\varphi_F$ 
instead of $\vpi$). 

In the previous section we have demonstrated the tight relations 
between $w_0$ and $\winf=-1+2/(9\al)$, $w_0$ and $w_a$, and 
hence $r$ and $w_0$, $w_a$. While $w_0$ and $w_a$ will eventually 
be determined by observations involving distances and 
the growth of cosmic structure (along with $r$ from CMB 
B-mode polarization, giving a strong 
consistency test), we already have bounds on dark energy 
indicating that its dynamics does not vary too much from 
cosmological constant behavior. For example, the reduced 
distance to CMB last scattering for $w_0=-0.8$ (and hence 
also with $w_a\approx-0.3$ by the thawing relation) differs from the 
LCDM prediction by 1.1\%, well above the Planck constraints 
\cite{planckR}. One would have to shift $\Om$ by 0.02 between 
the $(w_0,w_a)=(-0.8,-0.3)$ model -- giving a distance 
corresponding roughly 
to a constant $w=-0.9$ model -- and LCDM to move within 
the 0.4\% distance constraint. Thus we consider models more 
extreme, $w_0>-0.8$, to be disfavored. 

Since a strong connection exists between inflationary 
gravitational waves and dark energy dynamics in this 
model, then an experiment seeing a signature for one 
can inform the other type of experiment on the desired 
sensitivity level, i.e.\ ``where to look''. Furthermore, 
the thawing nature of the dark energy field raises the 
probative power of experiments that have good leverage 
on the dynamics in terms of $w_a$, as well as the recent 
universe value $w_0$.

%%%%%%%%%%%%%%%%%%%%%% 
\section{Conclusions} \label{sec:concl} 

We are on the cusp of experiments that will probe the 
early (inflation) and late (dark energy) epochs of 
cosmic acceleration with unprecedented accuracy. Lack of 
a signal of primordial gravitational waves and of dynamics 
distinct from a cosmological constant would give limited 
insight into the physics behind these fundamental phenomena. 
Quintessential inflation however ties these epochs together, 
and $\al$-attractors provide some definite predictions. 
The $\al$-attractor ExpLin model considered here has some 
important characteristics in terms of simple functions, 
symmetry protection against some quantum corrections, and 
avoidance of some fine tuning. 

The plateau with exponential potential part for inflation and 
the linear potential part (translating to an exponential potential 
free of a cosmological constant in the canonical field) for 
dark energy leads to tight relations between the inflation 
and dark energy observables. Late time acceleration arises 
out of a thawing field with a calibrated relation 
$w_a=-1.53(1+w_0)$ and dynamics leading to an attractor 
giving $1+w_0\approx0.5(1+\winf)$. Since $\winf$ is tied to 
the model parameter $\alpha$, as is the inflation tensor-to-scalar 
ratio $r$, this predicts strong connections between all the 
observables. We have verified this numerically and 
illustrated the insensitivity to initial conditions over a broad range. 

In a ``can't lose'' manner, values of $\al$ near the higher 
end of the fundamental physics predicted range will give a 
primordial gravitational wave signal, while those near the 
lower end will give a dark energy dynamics signal. The values 
near the middle of the range, e.g.\ Starobinsky inflation's 
$\al=1$, will give signals in both that should be accessible 
to next generation experiments -- an exciting prospect!

\acknowledgments 

This work was supported in part by the Energetic Cosmos 
Laboratory. EL is supported in part by the U.S.\ 
Department of Energy, Office of Science, Office of High Energy 
Physics, under contract number DE-AC02-05CH11231.

\bibliography{pgwde}

\end{document}